\documentclass[prl,twocolumn,superscriptaddress,floatfix,showpacs,letterpaper]{revtex4}
\usepackage[dvips]{graphicx}
\usepackage{amssymb}

\begin{document}

\title{Exact microscopic analysis of a thermal Brownian motor}

\date{\today}

\author{C. Van den Broeck}
\affiliation{Limburgs Universitair Centrum, B-3590 Diepenbeek,
Belgium}

\author{R. Kawai}
\affiliation{Department of Physics, University of Alabama at
Birmingham, Birmingham,  AL 35294}

\author{P. Meurs}
\affiliation{Limburgs Universitair Centrum, B-3590 Diepenbeek,
Belgium}

\begin{abstract}
We study a genuine Brownian motor by hard disk molecular
dynamics and calculate analytically its properties, including its
drift  speed and thermal conductivity, from microscopic theory.
\end{abstract}



\maketitle

It is (believed to be) impossible to systematically rectify
thermal fluctuations in a system at equilibrium. Such a perpetuum
mobile of the second kind, also referred to as a Maxwell
demon \cite{md}, would violate the second law of thermodynamics and
would, from the point of view of statistical mechanics, be in
contradiction with the property of detailed balance. Yet, it may
require quite subtle arguments to explain in detail on specific
models why rectification fails. Apart from the academic and
pedagogical interest of the question, the study of small scale
systems is motivated by rapidly increasing capabilities in
nanotechnology and by the huge interest in small scale
biological systems. Furthermore, when operating  under
nonequilibrium conditions, as is the case in living organisms, the
rectification of thermal fluctuations becomes possible. This
mechanism, also referred to as a Brownian motor \cite{rr}, could
furnish the engine that drives and controls the activity on a
small scale.
In this letter we propose a breakthrough in the theoretical and
numerical study of a small scale thermal engine. Our starting
point is the observation that one of the basic and most popular
models, namely the Smoluchowski-Feynman
ratchet \cite{rr,smoluchowski,feynman},
is needlessly complicated, and can be replaced by a simplified
construction involving exclusively hard core interactions.  Its
properties, including speed, diffusion coefficient and heat
conductivity, can be measured very accurately by hard disk
molecular dynamics and can be calculated exactly from microscopic
theory.

\begin{figure}[b]
\begin{center}
\includegraphics[width=3in]{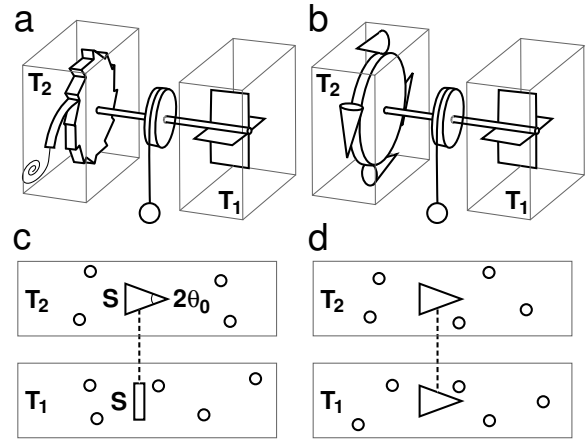}
\caption{ (a) Schematic representation of the Smoluchowski-Feynman
ratchet. (b) Similar construction without a pawl and spring. (c)
Two dimensional analogue referred to as the $AB$ motor.  The motor
is constrained to move along the horizontal $x$-direction (without
rotation or vertical displacement). The host gas consists of hard
disks whose centers collide elastically with the engine parts. The
shape of the arrow is determined by the apex angle $2\theta_0$ and
the vertical cross section $S$. Periodic boundary conditions are
used in the computer simulations. (d) A symmetric construction
referred to as the AA motor.} \label{fig:models}
\end{center}
\end{figure}

In Fig.~\ref{fig:models}a, we have schematically depicted the
construction originally introduced by Smoluchowski
\cite{smoluchowski} in his discussion of Maxwell demons and
re-introduced with two compartments at different temperatures by
Feynman \cite{feynman}. One compartment contains a ratchet with a
pawl and a spring, mimicking the rectifier device that is used in
clockworks of all kinds. The macroscopic mode of operation of such
an object generates the impression that only clockwise rotations
can take place, suggesting that this construction can be used as a
rectifier of the impulses generated by the impacts of the
particles in the other compartment on the blades with which the
ratchet is rigidly linked. As Feynman has argued, such a
rectification is only possible when the temperature in both
compartments is different.  We now introduce a model which is at
the same time a simplification and generalization of this
construction. First, we can dispose  of the pawl and spring in the
ratchet and consider any rigid but asymmetric object. An example
with a cone-shaped object in one compartment and a flat ``blade''
or  ``sail'' located in the other one is illustrated in
Fig.~\ref{fig:models}b. Second, we replace
 the single rotational degree of freedom with a single
translational degree of freedom (Figs.~\ref{fig:models}c and
\ref{fig:models}d). Third, we restrict ourselves to
two-dimensional systems. Finally, the substrate particles in the
various compartments are modeled by hard disks which undergo
perfectly elastic collisions with each other while their center
collides elastically with the edges of the motor \cite{remark}.

\begin{figure}[bt]
\begin{center}
\includegraphics[width=3in]{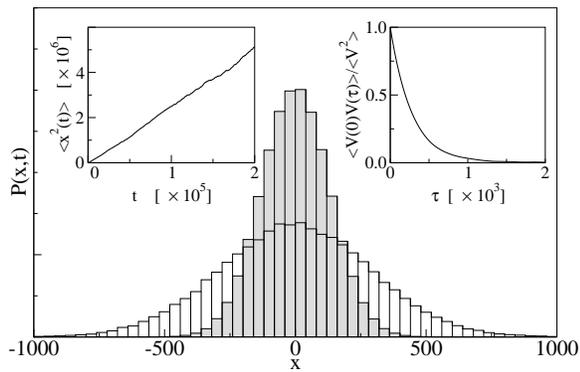}
\caption{Probability density $P(x,t)$ for the position $x$ of
motor $AB$ at times $t=1000$ (shaded) and $t=4000$ (open).
Inset left: mean square
displacement versus time. Inset right: velocity correlation
function.}
\label{fig:diffusion}
\end{center}
\end{figure}
\begin{figure}[bt]
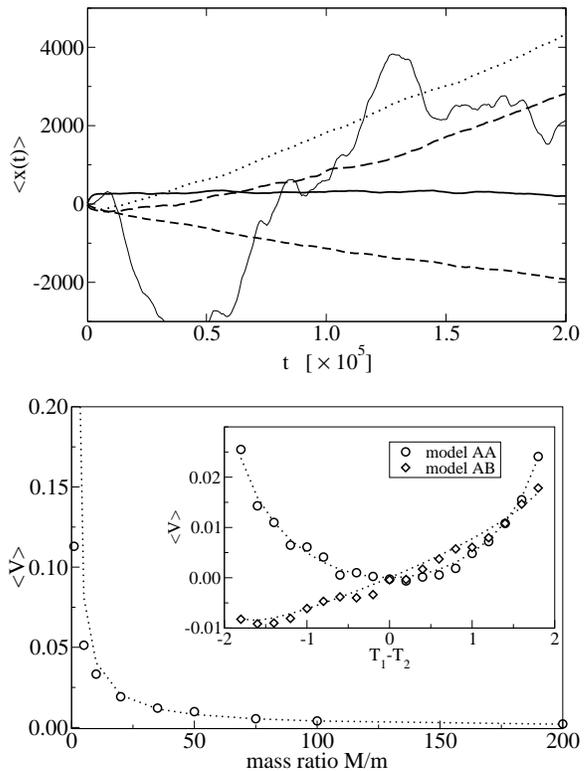

\begin{center}
\includegraphics[width=3in]{drift.eps}

\vspace{0.1in}

\centerline{\includegraphics[width=3in]{velocity.eps}}
\caption{Upper panel: Position of the motor as a function of time.
The thin solid curve shows a typical trajectory. All other curves
represent the average $\langle x(t) \rangle$.  The thick solid
line is the equilibrium case ($T_1=T_2=1$) for motor $AB$. The
dotted and dashed curves correspond to the nonequilibrium
situation for respectively motor $AA$ and $AB$ ($T_1=1.9$,
$T_2=0.1$). The situation with switched temperatures for $AB$ is
the dashed curve with a negative velocity. Lower panel: Average
velocity of motor $AB$ as a function of its mass $M$. Inset:
average speed (avg. over $2000$ runs) of motors $AA$ and $AB$ as a
function of the initial temperature difference $T_1-T_2$
($T_1+T_2=2$ fixed). The theoretical results (\ref{T.13b}) and
(\ref{T.13a}) predict lower speeds than the simulations.  However,
when the magnitude is scaled, the theoretical curves (dotted
lines) fit well with the simulations.} \label{fig:velocity}
\end{center}
\end{figure}
We first report on the results obtained from molecular dynamics
for two different realizations of our motor. The first one,
referred to as arrow/bar or $AB$ is inspired by the above
discussion. It consists of one triangular-shaped arrow in the
first compartment and a flat bar in the other
(Fig.~\ref{fig:models}c).  The other motor, called arrow/arrow or
$AA$, consists of an identical triangular-shaped arrow in both
compartments (Fig.~\ref{fig:models}d). Both units of the motor are
constrained to move as a rigid whole along the horizontal
$x$-direction as a result of their collision with the hard disks
in the two compartments. The initial state of the hard disk gases
corresponds to uniform (number) densities $\rho_1$ and $\rho_2$
and Maxwellian speeds at temperatures $T_1$ and $T_2$, in the
compartments $1$ and $2$ respectively, \cite{comment2} ($k_B=1$ by
choice of units). The boundary conditions are periodic both left
and right, and top and bottom. Unless mentioned otherwise, the
following parameter values are used: Each compartment is a $1200$
by $300$ rectangle and contains $800$ hard disks (mass $m=1$,
diameter $1$), i.e., particle densities $\rho_1=\rho_2=0.00222$.
Initial temperatures were set to $T_1=1.9$ and $T_2=0.1$. The
motor has a mass $M=20$, apex angle $2\theta_0=\pi/18$, and
vertical cross section $S=1$. The averages are taken over $1000$
runs.

When the temperatures are the same in both compartments,
$T_1=T_2$, no rectification takes place. In fact,
Fig.~\ref{fig:diffusion} shows that the motor undergoes plain
Brownian motion, with average zero speed, exponentially decaying
velocity correlations and linearly increasing mean square
displacement. The corresponding friction and diffusion coefficient
obey the Einstein relation. On the other hand, as soon as the
temperatures are no longer equal, the motor spontaneously develops
an average systematic drift along the x-axis.   The amplitude and
direction of the speed depend in an intricate way on the
parameters of the problem. In particular, the average speed
increases with the temperature  difference and the degree of
asymmetry (decreasing $\theta_0$) and decreases with increasing
mass of the motor roughly as $1/M$ (see the lower panel in
Fig.~\ref{fig:velocity}). Note furthermore that the observed
average speed can be very large, i.e. comparable to the thermal
speed $\sqrt{k_B T/M}$ of the motor. The $AA$ motor has a peculiar
behavior, resulting from the fact that both units are identical.
Whereas equilibrium is usually a point of flux reversal, the
velocity now displays a parabolic curve as a function of $T_1-T_2$
with a minimum equal to zero at the equilibrium state $T_1=T_2$.
It is clear from its symmetric construction that, at least when
$\rho_1=\rho_2$, an interchange of $T_1$ with $T_2$ can not modify
the speed so that the latter has to be an even function of
$T_1-T_2$, cf. Fig.~\ref{fig:velocity}.

\begin{figure}[bt]
\begin{center}
\includegraphics[width=3in]{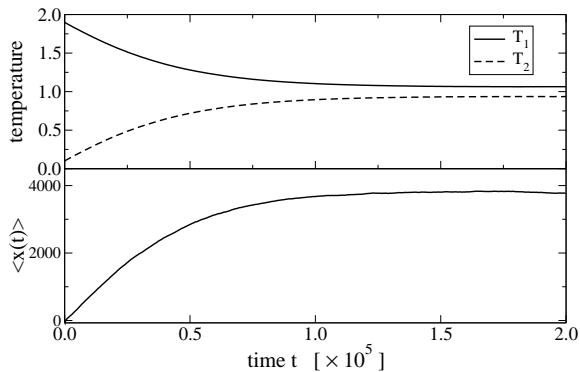}
\caption{Exponential decay of the temperatures to a final common
value ($T_{\text{final}}=(T_1+T_2)/2=1$.) in motor $AB$ and
concomitant disappearance of the average drift speed. To enlarge
the conductivity, a small mass $M=1$, a large vertical cross
section $S=10$, and a large apex angle $\theta_0=\pi/9$ are used.}
\label{fig:temp}
\end{center}
\end{figure}

We finally note that the observed systematic speed does not
persist forever. Indeed, the motion of the motor along the x
direction is a single degree of freedom that allows for
(microscopic) energy transfers hence thermal contact between the
compartments, a fact that was overlooked by Feynman in his
analysis and first pointed out in  \cite{parrondo,sekimoto}. As a
result one observes that the temperatures in both compartments
converge exponentially to a common final temperature with a
concomitant reduction and eventual disappearance of the systematic
motion as shown in Fig.~\ref{fig:temp}.  While this feature has
already been documented in detail in other constructions
\cite{vandenbroeck}, we have focused here on conditions in which
this conductivity is small and the compartments sufficiently large
so that one reaches a quasi-steady state with a well defined and
measurable average drift velocity.

To obtain analytic results from microscopic theory, which are
asymptotically exact, we focus  on the situation in which the
compartments are infinitely large while the  densities of the hard
disk gases are extremely low (more precisely, the so-called high
Knudsen number regime requires that  the mean free path is much
larger than the linear dimensions of the motor units). In this
limit, each compartment, characterized by its particle density
$\rho_i$ and temperature $T_i$, acts as an ideal thermal
reservoir.  We will furthermore assume that all the constituting
units of the motor are closed  and convex. Under these
circumstances, the motor never undergoes recollisions and the
assumption of molecular chaos becomes exact \cite{lk}. The
probability density $P(V,t)$ for the speed $\vec{V}=(V,0)$ of the
motor thus obeys the following Boltzmann-Master equation:
\begin{eqnarray}
&&\partial_t P(V,t) = \nonumber\\
&&\int dr \left[ W (V-r,r) P(V-r,t)-W(V,r) P(V,t)\right]
\label{T.0}
\end{eqnarray}
$W(V,r)$ is the transition probability per unit time for the motor
to change speed from $V$ to $V+r$ due to the collisions with the
gas particles in various compartments. The explicit expression for
$W(V,r)$  follows from  elementary arguments familiar from kinetic
theory of gases, taking into account the statistics of the
perfectly elastic collisions of the motor, constrained to move
along the $x$-direction, with the impinging particles:
\begin{widetext}
\begin{equation}
W(V,r)=\sum_i\int_0^{2\pi}d\theta
\int_{-\infty}^{\infty}dv_x\int_{-\infty}^{\infty}dv_y
\rho_i\phi_i(v_x, v_y) L_i F_i(\theta)
(\vec{V} -\vec{v}) \cdot \vec{e} (\theta) H[(\vec{V}-\vec{v})
\cdot\vec{e}(\theta)] \nonumber\\
\delta \left[r+\frac{m}{M} B(\theta)(V-v_x+v_y \cot\theta)\right]
\label{W}
\end{equation}
\end{widetext}
Here, the sum over $i$ runs over all the different compartments,
$L_i$ is the total circumference of the $i$th unit of the motor,
$B(\theta)=2M\sin^2{\theta} /(M+m\sin^2{\theta})$, $H[x]$ is the
Heaviside function, and $\vec{e}(\theta)=(\sin\theta,-\cos\theta)$
is the unit vector normal to a surface at angle  $\theta$, $\theta
\in [0,2\pi]$, the angles being measured counter-clockwise from
the horizontal axis. $\phi_i (v_x,
v_y)=m \exp{\left[-m(v_x^2 + v_y^2)/2k_BT_i\right]}/2\pi k_B T_i$ is the
Maxwellian velocity distribution in compartment $i$.
The shape of any closed convex unit of the motor is defined by the
(normalized) probability density $F(\theta)$ such that $F(\theta)
d\theta$ is the fraction of its outer surface that has an
orientation between $\theta$ and $\theta + d\theta$. Note that
$\langle\sin{\theta}\rangle=\langle\cos{\theta}\rangle=0$,
where the average is with respect to $F(\theta)$,
a property resulting from the
requirement that the object is closed.

The Boltzmann-Master equations (\ref{T.0}) and (\ref{W}) can now be
solved by a perturbation expansion in $\sqrt{{m}/{M}}$, following
a procedure similar to the one used in one-dimensional problems
such as the Rayleigh piston \cite{vankampen}  or the adiabatic
piston \cite{gruber}.  The details are somewhat involved and will
be given elsewhere \cite{meurs}.  To lowest order in the
perturbation, the Master equation  (\ref{T.0}) reduces to a
Fokker-Planck equation equivalent to the following linear Langevin
equation:
\begin{equation}
M\dot{V} =  -\sum\limits_i  \gamma_i V +
\sum\limits_i\sqrt{2\gamma_ik_B T_i} \;\eta_i \label{T.6}
\end{equation}
with  $\eta_i$ independent  Gaussian white noises of unit strength
and
\begin{equation}
\gamma_i = 4\rho_i L_i \sqrt{
\frac{k_B T_i m}{2\pi}} \int_0^{2\pi}  d\theta F_i(\theta)
\sin^2{\theta} \label{T.7}
\end{equation}
the friction coefficient experienced by the motor due to its
presence in compartment $i$. We conclude that at this order of the
perturbation the contributions from the separate compartments add
up and are each - taken separately - of the linear equilibrium
form. In particular the motor has no (steady state) drift
velocity, $\langle V \rangle = 0$.  It does however conduct heat.
In the case of two compartments $1$ and $2$, the heat flow per
unit time between them  is, as anticipated in Ref.
\onlinecite{parrondo}, given by a Fourier law:
$\dot{Q}_{1\rightarrow 2} = \kappa (T_1 - T_2)$ with conductivity
$\kappa={k_B} {\gamma_1\gamma_2}/[M({\gamma_1 + \gamma_2})]$. One
also concludes from (\ref{T.6}) and (\ref{T.7}) that the (steady
state) velocity distribution of the motor is Maxwellian, but at
the effective temperature
\begin{eqnarray}
T_{\text{eff}} = {\sum\limits_i \gamma_i T_i}\Big/{\sum\limits_i
\gamma_i}
\end{eqnarray}

At the next order of perturbation in $\sqrt{m/M}$, the
corresponding Langevin equation becomes  nonlinear in $V$ while at
the same time the Gaussian nature of the white noise is lost, a
feature well-known from the Van Kampen $1/\Omega$ expansion
\cite{vankampen}.
The most relevant observation is the
appearance, at the steady state, of a non-zero drift velocity:
\begin{eqnarray}
\langle V\rangle &=& \sqrt{ \frac{\pi k_B T_{\text{eff}}}{8M}}
\sqrt{\frac{m}{M}} \nonumber \\
&\times& \frac{\sum\limits_i \rho_i L_i \frac{T_i -
T_{\text{eff}}}{T_{\text{eff}}} \int_0^{2\pi}d\theta F_i (\theta)
\sin^3 \theta}{\sum\limits_i \rho_i L_i
\sqrt{\frac{T_i}{T_{\text{eff}}}} \int_0^{2\pi} d\theta
F_i(\theta) \sin^2(\theta)}\label{T.10}
\end{eqnarray}
This speed is of the order of the thermal speed of the motor,
times the expansion parameter $\sqrt{m/M}$, and further multiplied
by a factor that depends on the details of the construction. Note
that the Brownian motor ceases to function in the absence of a
temperature difference  ($T_i \equiv T_{\text{eff}},\forall i$)
and in the macroscopic limit $M\rightarrow \infty$ ($\langle V
\rangle \sim 1/M$). Note also that the speed is scale-independent,
i.e., independent of the actual size of the motor units: $\langle
V \rangle$ is invariant under the rescaling $L_i$ to $CL_i$.  To
isolate more clearly the effect of the asymmetry of the motor on
its speed, we focus on the case where the units have the same
shape in all compartments, i.e. $F_i(\theta) = F(\theta)$.  In
this case $T_{\text{eff}}$ is independent of $ F(\theta)$ and the
drift velocity is proportional to $\langle
\sin^3\theta\rangle/\langle \sin^2\theta\rangle$, with the average
defined with respect to $F(\theta)$.  The latter ratio is in
absolute value always smaller than $1$, a value that can be
reached for ``strongly'' asymmetric objects as will be shown below
on a particular example.

We now turn to a comparison between theory and simulations. From
the general result (\ref{T.10}), one obtains the following
expressions for the speed of the two motors that were studied:
\begin{eqnarray}
\langle V \rangle_{AA}
&=& \rho_1 \rho_2 (1-\sin \theta_0) \nonumber\\
&\times& \sqrt{\frac{m}{M}} \sqrt{\frac{\pi k_B}{8M}}
\frac{(T_1 - T_2) (\sqrt{T_1}-\sqrt{T_2})}{[\rho_1 \sqrt {T_1} +
\rho_2\sqrt{T_2}]^2}
\label{T.13b}
\end{eqnarray}
\begin{eqnarray}
\langle V \rangle_{AB}
&=& \rho_1 \rho_2 (1-\sin^2 \theta_0)
\sqrt{ \frac{m}{M}}
\sqrt{\frac{\pi k_B}{2M}}  \nonumber\\
&\times& \frac{(T_1 -
T_2) \sqrt{T_1}}{[2\rho_1 \sqrt {T_1} + \rho_2\sqrt{T_2} (1 + \sin
\theta_0)]^2} \label{T.13a}
\end{eqnarray}
In agreement with previous arguments, the AA motor, cf.
(\ref{T.13b}), always moves in the same direction, namely the
direction  of the arrow. Furthermore it is an example  where one
can increase the asymmetry to generate a maximum drift speed.  The
limit $|\langle \sin^3\theta\rangle| = \langle
\sin^2\theta\rangle$ is reached here when $\theta_0 \rightarrow
0$, which corresponds to an infinitely elongated and sharp arrow
in both compartments. Due to strong finite size effects (e.g.
sound waves among others), the agreement between the theoretical
results (\ref{T.13b}) and (\ref{T.13a}) and the computer
simulations is only qualitative: the theory predicts speeds which
are typically $20-40\%$ lower. However, Eqs.(\ref{T.13b}) and
(\ref{T.13a}) can be fitted to the simulation results by
appropriately rescaling the magnitude of the velocity (see Fig.
\ref{fig:velocity}), indicating that their dependencies on the
parameters, $M$, $T_1$ and $T_2$, are in good agreement with the
simulations.

In conclusion, we have provided a detailed analytic and numerical
study of a simplified version of the Smoluchowski-Feynman ratchet,
including an exorcism - based on microscopic theory - of its
operation as a Maxwell demon.

This work was supported in part by the National Science Foundation
under Grant No. DMS-0079478.


\begin{thebibliography}{99}
\bibitem{comment2}
Note that our system as a whole is finite and isolated so that
strictly speaking the equilibrium state should refer to a
micro-canonical ensemble.
\end{thebibliography}
\end{document}